\documentclass[aps,PRB,twocolumn,superscriptaddress,preprintnumbers]{revtex4-2}
\usepackage[T1]{fontenc}
\usepackage{times}
\usepackage{graphicx,color}
\usepackage{amsfonts,amsmath,amssymb,amsbsy}
\usepackage[colorlinks=true, linkcolor=blue, citecolor=blue, urlcolor=blue]{hyperref}
\usepackage{float}

\def\red#1{#1}

\begin{document}

\title{Experimental demonstration of position-controllable topological interface states\\
in high-frequency Kitaev topological integrated circuits}

\author{Tetsuya Iizuka}
 \affiliation{Systems Design Lab., School of Engineering, The University of Tokyo, Tokyo, Japan.
 \email{iizuka@vdec.u-tokyo.ac.jp}
 }%
\author{Haochen Yuan}
 \affiliation{Department of Electrical Engineering and Information Systems, The University of Tokyo, Tokyo, Japan.}%
\author{Yoshio Mita}
 \affiliation{Department of Electrical Engineering and Information Systems, The University of Tokyo, Tokyo, Japan.}%
\author{Akio Higo}
 \affiliation{Systems Design Lab., School of Engineering, The University of Tokyo, Tokyo, Japan.}%
\author{Shun Yasunaga}
 \affiliation{Department of Electrical Engineering and Information Systems, The University of Tokyo, Tokyo, Japan.}%
\author{Motohiko Ezawa}%
 \affiliation{Department of Applied Physics, The University of Tokyo, Tokyo, Japan.
 \email{ezawa@ap.t.u-tokyo.ac.jp}
 }%

\date{\today}
\begin{abstract}
\centerline{\textbf{abstract}} 
Topological integrated circuits are integrated-circuit realizations of topological systems.
Here we show an experimental demonstration by taking the case of the Kitaev topological superconductor model.
An integrated-circuit implementation enables us to realize high resonant frequency as high as 13GHz. 
We explicitly observe the spatial profile of a topological edge state.
In particular, the topological interface state between a topological segment and a trivial segment is the Majorana-like state.
We construct a switchable structure in the integrated circuit, which enables us to control the position of a Majorana-like interface state arbitrarily along a chain. Our results contribute to the development of topological electronics with high frequency integrated circuits.
\end{abstract}


\maketitle


\section{Introduction}\label{Sec:Intro}

Topological insulators and superconductors are fascinating new states of matter\cite{Hasan,Qi}. 
The Kitaev topological superconductor model\cite{Kitaev} is an intriguing one-dimensional (1D) systems realizing topological insulators and superconductors. Especially,  topological superconductors host Majorana edge states\cite{AliceaBraid,Been,Elli,Sato}, which are the key elements of a topological quantum computer\cite{KitaevQC,Nayak}.
The area of topological physics is expanded nowadays to photonic\cite{KhaniPhoto,Hafe2,Hafezi,WuHu,TopoPhoto,Ozawa}, acoustic\cite{Prodan,TopoAco,Berto,Xiao,He}, mechanical\cite{Lubensky,Chen,Nash,Paul,Sus,Sss,Huber,Mee} and electronic-circuit systems\cite{TECNature,ComPhys,Hel,Lu,Research,Zhao,YLi,EzawaTEC}. They are called artificial topological systems. There are several merits which are difficult to be achieved in inorganic crystals: 1) It is possible to make a fine tuning of the system, which is crucial for observing topological edge states. 2) It is possible to construct a few site systems. 3) It is possible to directly measure the site dependent information.

It is a nontrivial task to materialize the Kitaev topological superconductor model because it involves complex hoppings. 
It describes a \textit{p}-wave  topological superconductor, 
although the Majorana edge state itself can be generated in a \textit{s}-wave superconductor  with the aid of a topological insulator nanowire\cite{Mourik,Agha}. We note that there is no physical realization of the Kitaev topological superconductor model so far.

Electronic circuits present an ideal platform to realize various topological phases%
\cite{TECNature,ComPhys,Hel,Lu,YLi,EzawaTEC,Research,Zhao,EzawaLCR,EzawaSkin,Garcia,Hofmann,EzawaMajo,Tjunc,Lee}. 
The emergence of topological edge states is observed by means of impedance resonance.
However, experimental demonstrations have so far been restricted mainly to printed circuit boards with discrete components, 
except for a simulation of the Su-Schrieffer-Heeger model\cite{IC} and  
an integrated-circuit realization of Floquet's topological insulator\cite{Floquet}.
The integrated-circuit realization is an important step toward industrial applications of topological electronics.

In order to generate Majorana-like states, it is necessary to simulate electron and hole bands in electronic circuits.
Although there is a theoretical proposal with the use of chains of capacitors and inductors\cite{EzawaMajo,Tjunc}, 
there is so far no experimental demonstration of this theoretical proposal. 

Most of previous experiments were carried out based on patterned structures, where it is impossible to control the topological and trivial phases once the sample is manufactured. Actually, it is very hard to introduce switch structures in inorganic materials, photonic crystals and acoustic systems. On the other hand, transistors act as switches in electronic circuits and hence, there is a possibility to construct a switchable topological system based on electronic circuits.

In this paper, we perform an experimental demonstration of switchable topological integrated circuits, which are integrated-circuit realizations of topological systems, by taking the case of the Kitaev model.
An integrated-circuit implementation enables us to realize high resonant frequency. In this paper, we realized a Kitaev chain implementation whose resonant frequency is as high as 13GHz. 
We explicitly observe the spatial profile of a topological edge state and determine its penetration length.
The system may contain several topological and trivial segments simultaneously along a chain.
In particular, we observe the signal of a Majorana-like state emerging at the interface of a topological segment and a trivial segment.
It is topologically protected since it necessarily emerges between the topological and trivial segments.
These two topologically different segments are interchangeable simply by switching between inductors and capacitors.

\section{Results}\label{Sec:Results}

\begin{figure*}[t]
\centering
\includegraphics[width=0.98\linewidth]{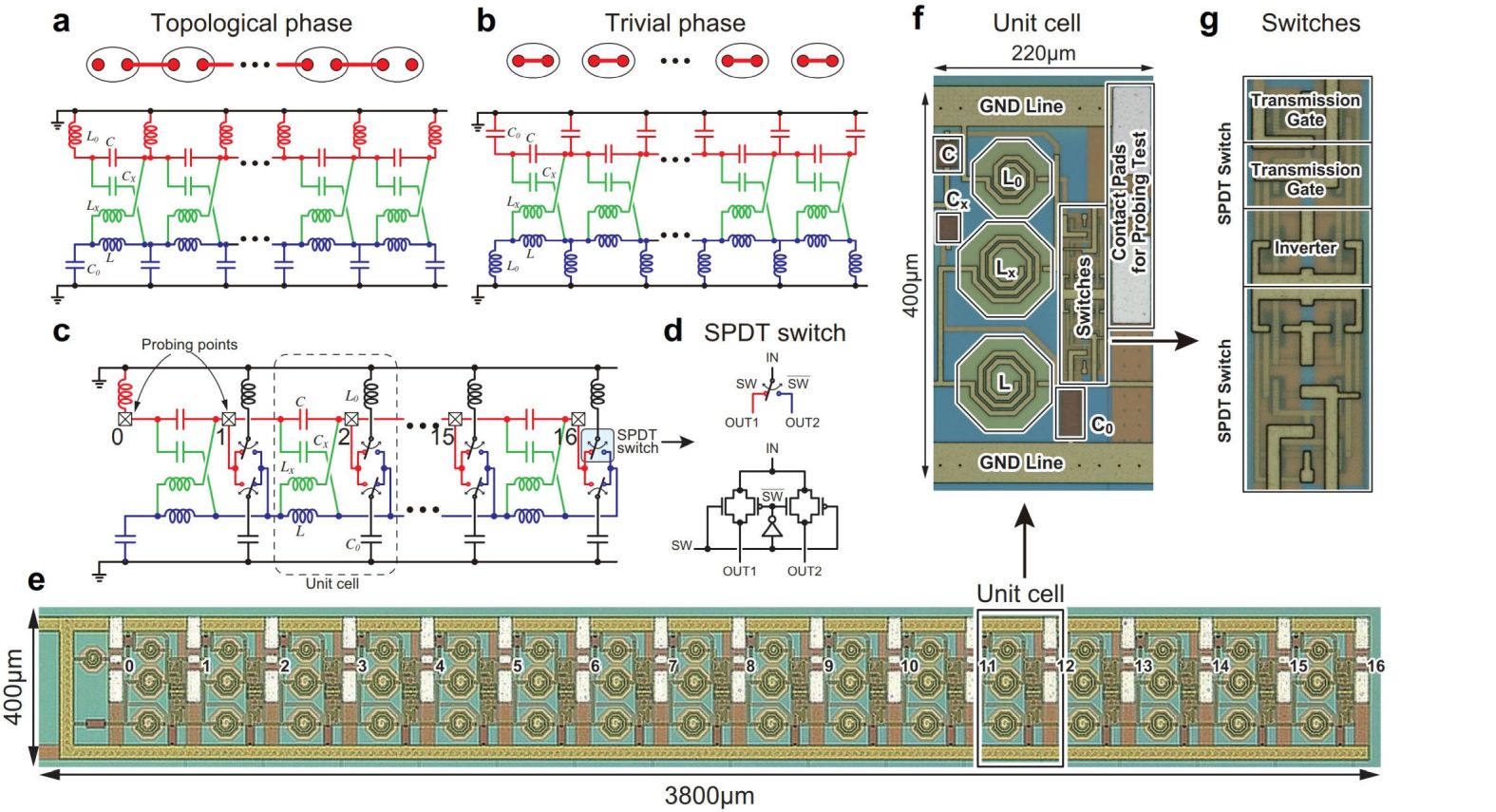}
\caption{\textbf{Kitaev chain.} \textbf{a}, \textbf{b}  and \textbf{c}, The electronic-circuit representation of the Kitaev chain. 
\textbf{a}, All-topological configuration where the topological edge state emerges at both the left and right edges of the chain. 
\textbf{b}, All-trivial configuration that does not have a topological edge state. 
\textbf{c}, The implemented state-configurable Kitaev chain circuit. 
\textbf{d}, By using two single-pole double-throw (SPDT) switches with inverters in the unit cell, the connection of $L_0$ and $C_0$ can be swapped to change its topological/trivial state. The SPDT switch is realized by two Complementary Metal-Oxide-Semiconductor (CMOS) transmission gate switches.
\textbf{e}, A picture of an 16-unit cell integrated circuit for the Kitaev chain.
\textbf{f}, A picture of a unit cell.
\textbf{g}, A zoom of SPDT switches in Fig.\ref{fig:KitaevCirc}\textbf{f}.
Each SPDT switch is composed of an inverter and two transmission gates with n-type and p-type Metal-Oxide-Semiconductor field-effect transistors as in Fig.\ref{fig:KitaevCirc}\textbf{d}.
}
\label{fig:KitaevCirc}
\end{figure*}

\begin{figure*}[t]
\centering
\includegraphics[width=0.98\linewidth]{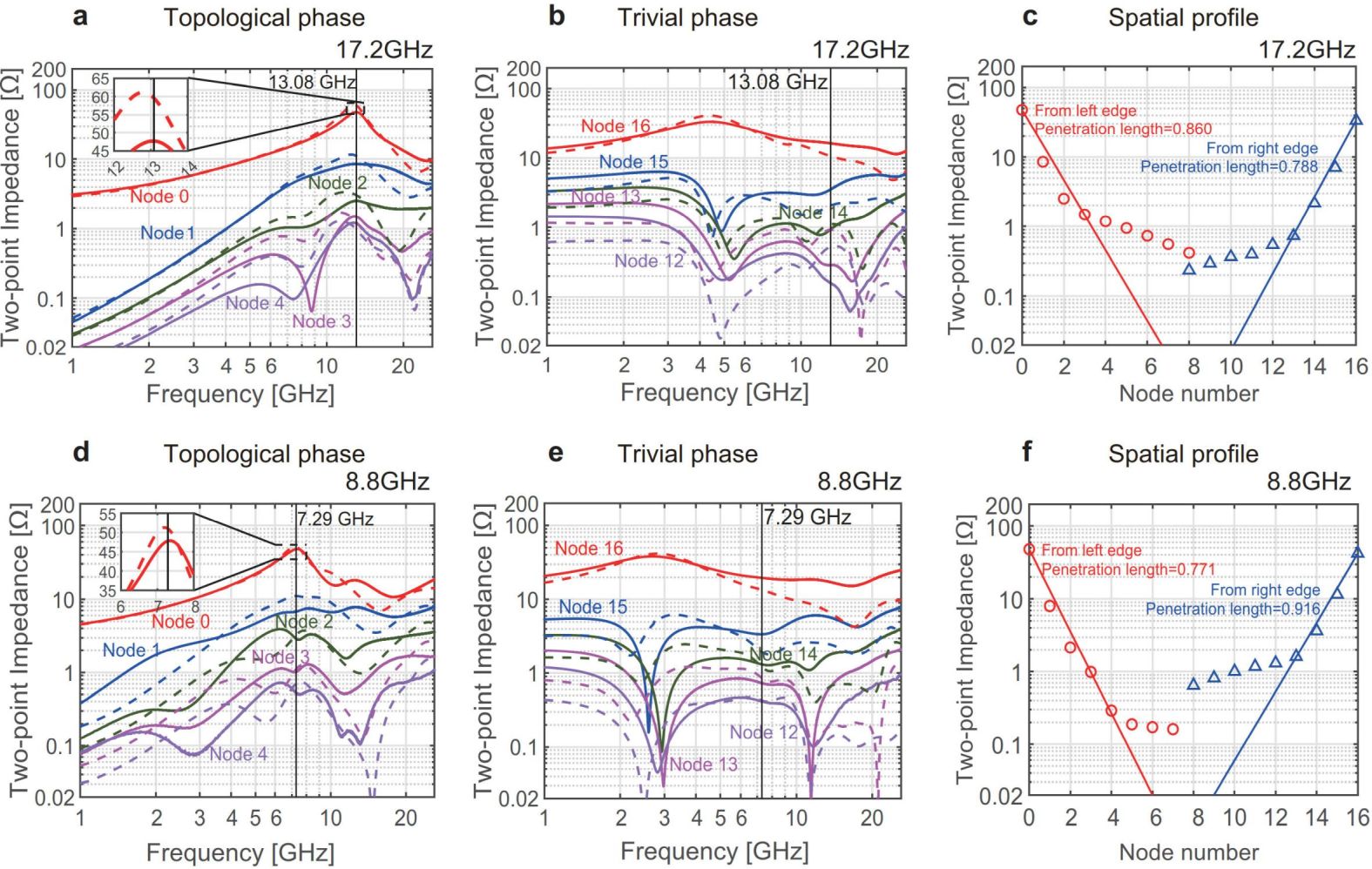}
\caption{\textbf{Impedance measurements.}
\textbf{a}, Frequency dependence of the impedance measured from the left edge of the electronic circuit with the characteristic frequency $\omega_{\text{resonant}}=$13.1\,GHz Kitaev chain for all-topological setup.  
\textbf{b}, Frequency dependence of the impedance measured from the right edge of the electronic circuit with the characteristic frequency $\omega_{\text{resonant}}=$13.1\,GHz Kitaev chain for all-trivial setup.  
Solid and dashed lines show the measured and simulated results of the Kitaev chain, respectively.
\textbf{c}, The spatial profile of the impedance values for all-topological mode measured from both left and right edges at their measured resonant frequencies. 
\textbf{d}, \textbf{e} and \textbf{f}, Similar results for the circuit with the characteristic frequency $\omega_{\text{resonant}}=$7.3\,GHz.
The solid and dashed curves show measurement and simulation results, respectively.}
\label{fig:Results}
\end{figure*}

\smallskip
\subsection{Kitaev chain}

The Kitaev chain model is the basic model of a topological superconductor. 
Our main result is its implementation in an integrated electronic circuit.
To realize a Cooper pair it is necessary to incorporate  an electron band and a hole band together with cross terms between these two bands into the circuit, as shown in Methods.

We first illustrate an electronic circuit for the Kitaev chain~\cite{EzawaMajo,Tjunc} in Fig.~\ref{fig:KitaevCirc}a, b and c. The capacitor channel (indicated in red) corresponds to the electron band, while the inductor channel (in blue) corresponds to the hole band. The two main channels are crosslinked through $C_x$ and $L_x$. Each node is connected to the ground via an inductor $L_0$ or a capacitor $C_0$ to realize a topological state or a trivial state, respectively, as shown in Figs.~\ref{fig:KitaevCirc}a and b.
The topological phase is realized by the configuration shown in Fig.~\ref{fig:KitaevCirc}a, while the trivial phase is realized by the configuration shown in Fig.~\ref{fig:KitaevCirc}b. 

A single Kitaev chain may accommodate several segments which are either topological or trivial.
A Majorana-like state emerges at an interface between the two phases. 
We introduce two single-pole double-throw (SPDT) switches in each unit cell as illustrated in Fig.~\ref{fig:KitaevCirc}c. The electric circuit for the SPDT switch is shown in Fig.~\ref{fig:KitaevCirc}d. 
The switching is done by swapping the connection of $L_0$ and $C_0$, 
by way of which the position of a Kitaev interface state is controlled.
In the integrated circuits, the SPDT switch is simply implemented with an inverter and two Complementary Metal-Oxide-Semiconductor (CMOS) transmission gates, composed of n-type and p-type metal oxide semiconductor field-effect transistors as shown in Fig.~\ref{fig:KitaevCirc}f and g.

The Kitaev chain circuit shown in Fig.~\ref{fig:KitaevCirc}c is implemented onto the chip using 180\,nm CMOS technology as shown in Fig.~\ref{fig:KitaevCirc}e. On a 5\,mm$\times$5\,mm chip, two 16-unit cell Kitaev chain circuits were integrated for two different target resonant frequencies, 7.3\,GHz and 13.1\,GHz. We show a zoom-in view of the unit cell layout in Fig.~\ref{fig:KitaevCirc}f, which shows that it includes 3 inductors $L$, $L_x$ and $L_0$, 3 capacitors $C$, $C_x$ and $C_0$, 2 SPDT switches, and a contact pad at each node for direct probing measurement with GSG (Ground, Signal, Ground) probes.  A photo of the SPDT switches is shown in Fig.~\ref{fig:KitaevCirc}g. Two transmission gates and an inverter are integrated for each SPDT switch. 
The values for the capacitors and inductors are summarized in TABLE~\ref{tab:parameters}.

\smallskip
\subsection{Topological edge states}

Fig.~\ref{fig:Results}a, b and c summarizes the impedance measurement results of the Kitaev chain designed for 13.1\,GHz resonant frequency. 
Figs.~\ref{fig:Results}a and b show the frequency dependence of the impedance measured from the left edge of the chain for topological setup and right edge of the chain for trivial setup, respectively. The solid and dashed lines show measurement and simulation results, respectively. 

\begin{table}
\centering
\caption{\textbf{Parameters used for the Kitaev chain.} $C$, $C_x$ and $C_0$ represent the capacitance, while $L$, $L_x$ and $L_0$ represent the inductance.}\label{tab:parameters}
\begin{tabular}{c|c|c} \hline
& 7.3\,GHz & 13.1\,GHz \\ \hline
$C$ & 440\,fF & 220\,fF \\
$L$ & 747\,pH & 384\,pH \\
$C_x$ & 396\,fF & 204\,fF \\
$L_x$ & 830\,pH & 427\,pH \\
$C_0$ & 880\,fF & 440\,fF \\
$L_0$ & 374\,pH & 192\,pH \\ \hline
\end{tabular}
\end{table}

\begin{figure*}[t]
\centering
\includegraphics[width=0.6\linewidth]{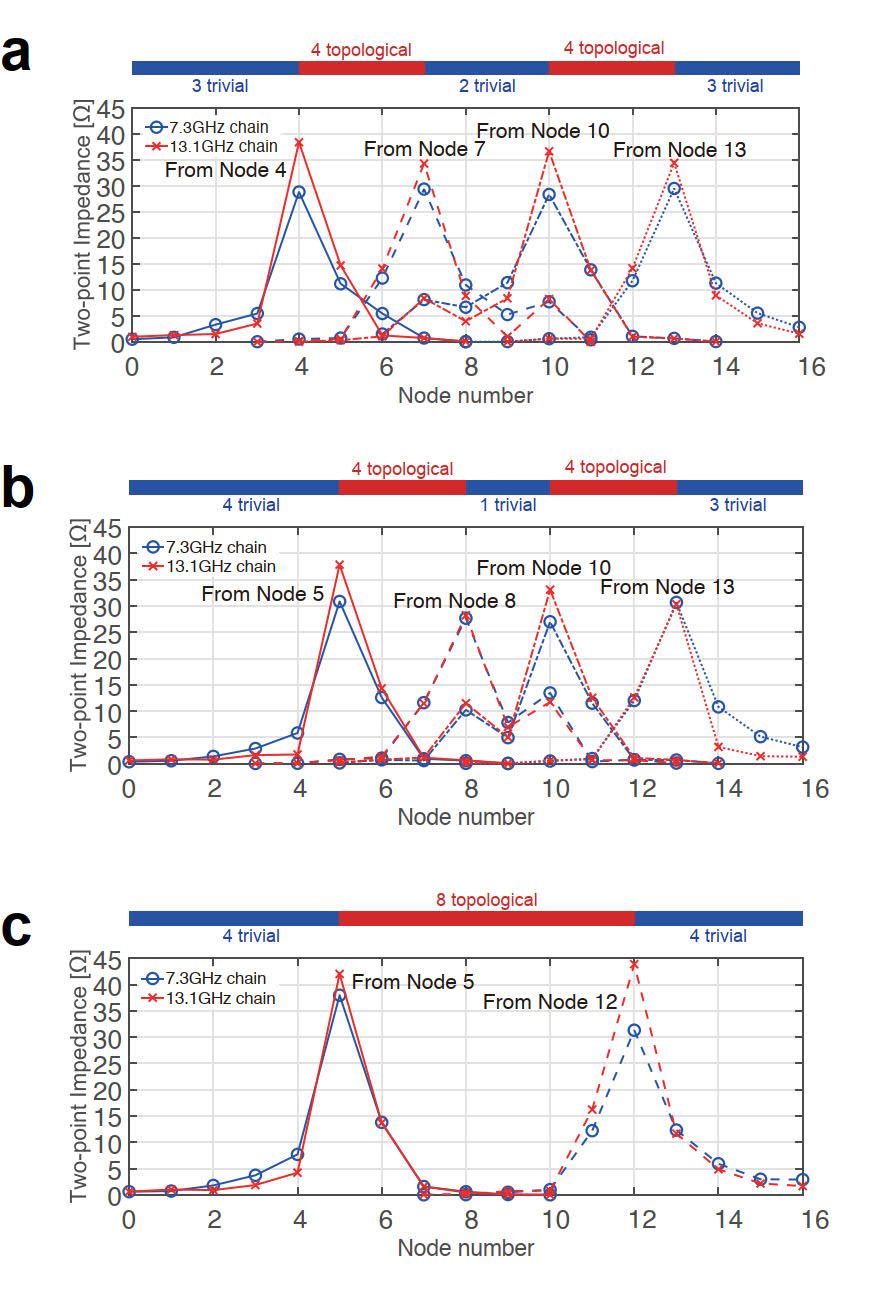}
\caption{\textbf{Topological interface states.} Measurement results of the topological edge state locations depending on different Kitaev chain configurations. 
"$n$ trivial (topological)" indicates that the trivial (topological) segment contains $n$ unit cells.
Blue (red) data points are for 7.3\,GHz (13.1\,GHz) chain.
\textbf{a}, The topological edge states emerge at 4th, 7th, 10th and 11th nodes. \textbf{b}, The locations of the edge states move to the 5th, 8th, 10th and 11th nodes. \textbf{c}, When two topological segments combine to one segment, the edge states emerge only at 5th and 12th nodes.}
\label{fig:edgestate}
\end{figure*}

\begin{figure*}[t]
\centerline{\includegraphics[width=0.88\textwidth]{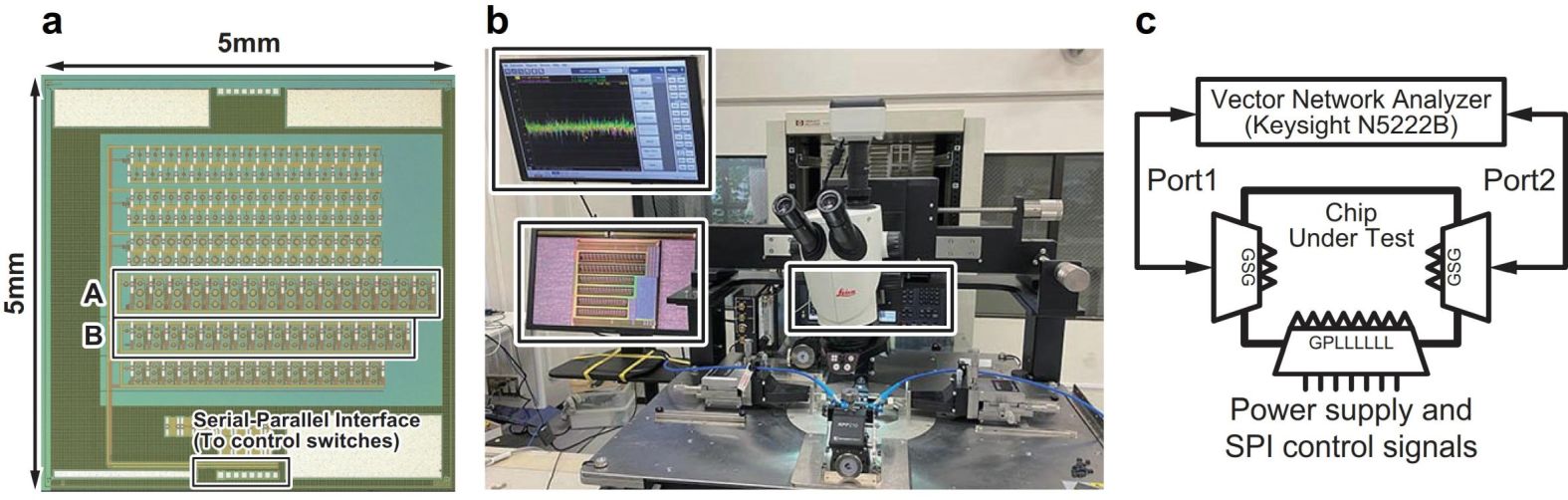}}
\caption{\textbf{Setup.} \textbf{a}, A microphotograph of the chip for the Kitaev chain,
where A (B) shows the circuits for the  16-stage Kitaev model with 7.3\,GHz~(13.1\,GHz).
 \textbf{b}, A photo of the measurement setup 
 \textbf{c}, A block diagram of the measurement setup including power supply and
serial-parallel interface (SPI) control signals.}
\label{FigSetup}
\end{figure*}

As we can see from the impedance peak of the leftmost edge in Fig.~\ref{fig:Results}a, the measured resonant frequency is 13.08\,GHz, while the resonant frequency directly calculated based on the on-chip $L$ and $C$ values in Table~\ref{tab:parameters} is 17.2\,GHz.
This frequency shift is mainly caused by the parasitic inductance of the metal wires in the unit cell to connect the circuit elements as well as the parasitic capacitance introduced by the SPDT switches realized with transistors. Without considering the wires and transistors, the simulated resonant frequency is 16.4\,GHz, which is closer to the theoretical value. Since the parasitic impact is inevitable on the integrated chip, we utilized a detailed electro-magnetic simulation to tune the actual resonant frequency. 
In addition, for both the topological and trivial setups, the measurement results show discrepancy from the simulation results mainly due to the imperfect transistor model. 
Especially, the peaking characteristics becomes less obvious in the topological setup. To verify the impact of the switch transistors, we have also designed and measured the Kitaev chain without switches as shown in Supplementary Fig.1. When the chains are composed only of the passive components such as inductors and capacitors, the measurement results agree almost perfectly with the simulation results. See Supplementary Note 1 for more details, where measurement data is shown in Supplementary Fig.2.

Fig.~\ref{fig:Results}b shows that no impedance peaks are observed at the resonant frequency in the trivial phase.

Fig.~\ref{fig:Results}c summarizes the two-point impedance values at their measured resonant frequencies  when all the system is in the topological phase.
The blue and red lines show the impedance measured from the left and the right edges, respectively. 
The leftmost (0-th) and rightmost (16-th) node impedance correspond to $Z_{11}$ value of the $2\times2$ impedance matrix. 

In the topological setup, the impedance peaks are observed at both the edges. The penetration length of the topological edge state is 0.860 unit cell for the left edge and 0.788 unit cell for the right edge. The discrepancy from the theoretical value 0.610 unit cell is mainly caused by the SPDT switches designed with transistors. See Supplementary Note 2 for details. A possible reason would be the hybridization
effect of two topological edge states for a finite length chain. However, this is not the case. See Supplementary Note 3 and Supplementary Fig.3.

We have also carried out a measurement for the Kitaev chain designed for 7.3\,GHz, whose results are shown in Fig.~\ref{fig:Results}d, e and f.
As we can see from the impedance peak of the leftmost edge in Fig.~\ref{fig:Results}d, the measured resonant frequency is 7.29\,GHz, while the resonant frequency directly calculated based on the on-chip $L$ and $C$ values in Table~\ref{tab:parameters} is 8.8\,GHz. The simulated resonant frequency without wire is 8.6\,GHz, which is much closer to the calculated value. Again, the shift in the resonant frequency is due to the effect of the parasitics which is inevitable on the integrated chip. For precise estimation of the actual resonant frequency including the impact of wirings, we utilized the EM simulation.
In the topological setup, the impedance peaks are observed at both the edges. The penetration length of the topological edge state is 0.771 unit cell for the left edge and 0.916 unit cell for the right edge, while  the theoretical value is 0.680 unit cell.

\smallskip
\subsection{Topological interface states}

We have so far observed the topological edge states. There is also a topological interface state between topological and trivial phases. It is possible to switch the topological and trivial phases for each segment. Fig.~\ref{fig:edgestate} summarizes the 2-point impedance 
at the resonant frequency with 3 different
switch configurations for the Kitaev chains with 7.3\,GHz and
13.1\,GHz designs. In Fig. 3a we divided the chain into 4 segments.
 The impedance peak that corresponds to the topological interface state emerges at the edges of the topological segments. When we move the left topological segment 
to the right by one unit, the location of the edge states moves accordingly as shown in Fig.~\ref{fig:edgestate}b. Then if the two separated topological segments are combined into one segment as shown in Fig.~\ref{fig:edgestate}c, we observe only two impedance peaks at the left and right edges of the single topological segments. This clearly demonstrates the movement of the topological interface state that emerges on the electronic-circuit realization of the Kitaev chain implemented onto the integrated circuit. We also observe the same behavior for two chains with different resonant frequencies, which proves that the topological interface state emerges independent of the designed resonant frequency.

\smallskip
\section{Conclusion}

We have materialized the Kitaev model in integrated circuits. 
The model has topological and trivial phases.
It is possible to create several segments which are either topological or trivial in a single chain.
Topological edge states emerge at both the edges of a topological segment, which are observable by mean of the impedance resonance. 

We have demonstrated that the segment size can be as small as one unit cell because the penetration length can be made smaller than one unit cell: See Fig.\ref{fig:edgestate}b.
Furthermore, we have equipped our integrated circuit with a switchable structure, which enables us to control the position of a topological interface state arbitrarily along a chain. 
Such a possibility is a great merit of topological electric circuits over other artificial topological systems, where an integrated topological pattern is printed once and for all.

We have observed that the resonant frequency is lower than the theoretical value estimated from $\omega_{\text{resonant}}=1/\sqrt{LC}$. This is due to the parasitic inductance present in the wires. Details are shown in Supplementary Note 4 and Supplementary Fig.4.

The integrated circuit has small inductance and capacitance, which leads to high frequency operation. Indeed, the characteristic frequency is 13GHz. It is much larger than the previous integrated circuit implementation\cite{Floquet}, where the characteristic frequency is 730MHz.
The size of the unit cell is 200$\mu$m and hence, largely integrated circuits are possible. 
The sample randomness is of the order of 1$\%$ in our sample. It is much smaller than that of commercially obtained inductors, where the randomness is of the order of 10$\%$. The preciseness of circuit elements is beneficial for a sharp topological peak. Actually, particle-hole symmetry is slightly broken in our electric circuit due to the randomness. However, the topological peak is experimentally well observed because the topological peak is robust with 1$\%$ randomness as shown in Supplementary Fig.5 in Supplementary Note 5.
Mass production is possible in integrated circuits, which will benefit for future industrial applications of topological electronics.

In \cite{Floquet}, Floquet's topological insulator is implemented onto the integrated circuit chip with 40\,nm technology, which is applied to a wireless communication with a 4-element phased array antenna using 730\,MHz carrier frequency. Since our 1-dimensional Kitaev chain enables the impedance switching by changing the position of the interface between topological and trivial segments, this structure can be applied as a high-frequency path selectors or switches based on impedance matching. Especially, our implementation achieves the resonant frequency more than 10\,GHz and can be extended to higher frequency bands. Our results will be applicable to future 5G technology as in the case of the previous study\cite{Floquet}.


\smallskip

\section{Methods}

\subsection{Measurements}
A block diagram and a photo of the measurement setup are shown in Fig.\ref{FigSetup}. We observed the topological edge state based on two-point impedance measurement.

We observe two-point impedance with a vector network analyzer (VNA), Keysight N5222B. The chip measurement is done on the probe station, Formfactor Summit11000. A 2$\times$2 Z-matrix is derived from the 2$\times$2 S-parameter measured by the VNA. 
The chain configuration (the state of the SPDT switches) is controlled by the serial-parallel interface (SPI) integrated on the same chip, whose configuration data are written from an external PC.

Simulation is done with a circuit simulator, Cadence Spectre. The S-parameters of the passive components such as capacitors and inductors are extracted for circuit simulation with Cadence EMX, which is a planar 3D electromagnetic simulator based on the Fast Multipole Method (FMM) designed for high-frequency integrated circuits.

\subsection{1D $p$-wave Kitaev topological superconductor model}
The original Kitaev $p$-wave superconductor model is defined on the 1D lattice as
\begin{eqnarray}
H=-\mu \sum_{x}c_{x}^{\dagger }c_{x}-\frac{t}{2}\sum_{x}\left(
c_{x}^{\dagger }c_{x+1}+c_{x+1}^{\dagger }c_{x}\right) \notag\\
-\frac{1}{2}%
\sum_{x}\left( \Delta e^{i\phi }c_{x}c_{x+1}+\Delta e^{-i\phi
}c_{x+1}^{\dagger }c_{x}^{\dagger }\right) ,  \label{KitaevPwaveHamil}
\end{eqnarray}%
where $\mu $ is the chemical potential, $t>0$ is the nearest-neighbor
hopping strength and $\Delta >0$ is the $p$-wave pairing amplitude of the
superconductor. 

By introducing the Nambu representation $\Psi _{k}^{\dagger }=\left(
c_{k}^{\dagger },c_{-k}\right) $ and $\Psi _{k}=\left( c_{k},c_{-k}^{\dagger
}\right) ^{T}$ one can write the Hamiltonian in the Bogoliubov-de Gennes form
\begin{equation}
H=\frac{1}{2}\sum_{k}\Psi _{k}^{\dagger }H(k) \Psi _{k},  \label{BdG}
\end{equation}
with a 2$\times$2 form Hamiltonian
\begin{equation}
H(k)=\frac{1}{2}\left(\begin{array}{cc}
-t \cos k-\mu & i \Delta_0 \sin k \\
-i \Delta_0 \sin k & t \cos k+\mu
\end{array}\right).
\end{equation}
The zero-energy state of the Bogoliubov-de Gennes Hamiltonian is a Majorana state, and hence, there appear Majorana edge states in the topological phase of the Kitaev model.

Here, $t, \mu, \sigma_i$ and $\Delta_i$ represent the hopping amplitude, the chemical potential, the spin degree of freedom, and the superconducting gap parameter, respectively. 
It is well known that the system is topological for $|\mu|<|2 t|$ and trivial for $|\mu|>|2 t|$ irrespective of $\Delta_i$ provided $\Delta_i \neq 0$.

We then realize this $p$-wave Kitaev model by way of an electronic circuit. As shown in Fig.\ref{fig:KitaevCirc}a, this circuit chain contains two main lines, one connected by a series of capacitors $C$ implementing the electrons band, while another connected by a series of inductors $L$ implementing the holes band, respectively. Pairing interaction between the two bands is simulated by bridging capacitors $C_x$ and inductors $L_x$.  Each electron node and each hole node is connected to the ground via a capacitor $C_0$ and inductors $L_0$, respectively. The hopping amplitudes $t$ realized in the electrons band and holes band are opposite since the capacitors $C$ contained in the electrons band contribute the terms $i \omega C$ while the inductors $L$ contained in the holes band contribute the terms $1 /(i \omega L)$. 

The circuit Laplacian is given by
\begin{equation}
J_{a b}(\omega)=\left(\begin{array}{ll}
f_1 & g_1 \\
g_2 & f_2
\end{array}\right),
\end{equation}
where
\begin{equation}
\begin{aligned}
&f_1=-2 C \cos k+2 C-\left(\omega^2 L_0\right)^{-1} \\
&f_2=2\left(\omega^2 L\right)^{-1} \cos k-2\left(\omega^2 L\right)^{-1}+C_0 \\
&g_1=-C_x e^{i k}+\left(\omega^2 L_x\right)^{-1} e^{-i k} \\
&g_2=\left(\omega^2 L_x\right)^{-1} e^{i k}-C_x e^{-i k},
\end{aligned}
\end{equation}
for topological phase and 
\begin{equation}
\begin{aligned}
&f_1=-2 C \cos k+2 C+C_0\\
&f_2=2\left(\omega^2 L\right)^{-1} \cos k-2\left(\omega^2 L\right)^{-1} -\left(\omega^2 L_0\right)^{-1} \\
&g_1=-C_x e^{i k}+\left(\omega^2 L_x\right)^{-1} e^{-i k} \\
&g_2=\left(\omega^2 L_x\right)^{-1} e^{i k}-C_x e^{-i k},
\end{aligned}
\end{equation}
for trivial phase.

The essence to realize the 1D model in circuit form is to make the circuit Laplacian equal to the system Hamiltonian. Clearly, to make it possible, particle-hole symmetry (PHS) must be respected, which requires these three pairs of LC resonators shares the same resonant frequency, that is, 
\begin{equation}
\omega_{\text{resonant}} \equiv 1 / \sqrt{L C}=1 / \sqrt{L_0 C_0}=1 / \sqrt{L_x C_x}. 
\end{equation}
Once PHS is respected, the relationship between circuit components and Hamiltonian parameters could be induced and expressed as follows:
\begin{equation}
\left\{\begin{array}{l}
t=-C, \\
\mu=-2 C+C_0, \\
\Delta_0=-C_x.
\end{array}\right.
\end{equation}

To make the 1D circuit chain topological, we set $\mu$ to 0 to meet the topological mode requirements of $|\mu|<|2 t|$. This topological property is satisfied by the emergence of grounded capacitors $C_0$ and inductors $L_0$, since the system will be precisely located at the critical point between the topological and trivial states. Therefore, by exchanging the connections of $C_0$ and $L_0$, we could perform transitions between these two states.

\subsection{Impedance resonance}
The emergence of a topological edge states is observed via impedance resonance. The topological edge state is a zero-energy
eigenstate of the Hamiltonian. It corresponds to the zero
admittance, and hence, the emergence is observable by the divergence in the impedance. 

The two-point impedance between the $a$ and $b$ nodes is given by\cite{Hel} 
\begin{equation}
Z_{ab}\equiv V_{a}/I_{b}=G_{ab}, 
\end{equation}
where $G$ is the Green
function defined by the inverse of the Laplacian $J$, $G\equiv J^{-1}$, $V_{a}$ is the voltage at  site $a$ and $I_b$ is the current at  site $b$.

\textbf{Data Availability.} 

The data that support the findings of this study are available from the corresponding
author upon reasonable request.

\newpage

\nocite{*}


\small

\smallskip
{\normalsize{\textbf{Acknowledgments}}}\\
This work is supported by CREST, JST (Grants No. JPMJCR20T2). The authors would like to thank Z.~Yang, S.~Li and X.~Chen for their measurement support. The on-chip passive components are designed based on RF cell library developed by Umeda laboratory, Tokyo University of Science. The LSI chip in this study was designed and fabricated through the activities of VDEC, The University of Tokyo, in collaboration with Cadence Design Systems, Rohm Corporation and Toppan Printing Corporation.

\smallskip
{\normalsize{\textbf{Author contributions}}}\\
M.E., Y.M. and T.I. planned the study. 
T.I. and H.Y. designed the topological circuits and performed the experiments.
T.I., M.E. and H.Y. collected and analyzed the data.
M.E. and T.I. wrote the manuscript with input from H.Y., Y.M., A.H. and S.Y.
All the authors discussed the project and the results.

\smallskip
{\normalsize{\textbf{Additional information}}}\\
Supplementary information is available.

\smallskip
{\normalsize{\textbf{Competing financial and non-financial interests}}}\\
The authors declare no competing interests.

\clearpage\newpage
\onecolumngrid
\def\theequation{S\arabic{equation}}
\def\thefigure{S\arabic{figure}}
\def\thesubsection{S\arabic{subsection}}
\setcounter{figure}{0}
\setcounter{equation}{0}
\setcounter{section}{0}

\centerline{\textbf{\Large {Supplementary Information}}}
\bigskip
\bigskip

\section{Supplementary Note 1. Measurement results of Kitaev model without switches}

\begin{figure}[h]
\centerline{\includegraphics[width=0.65\textwidth]{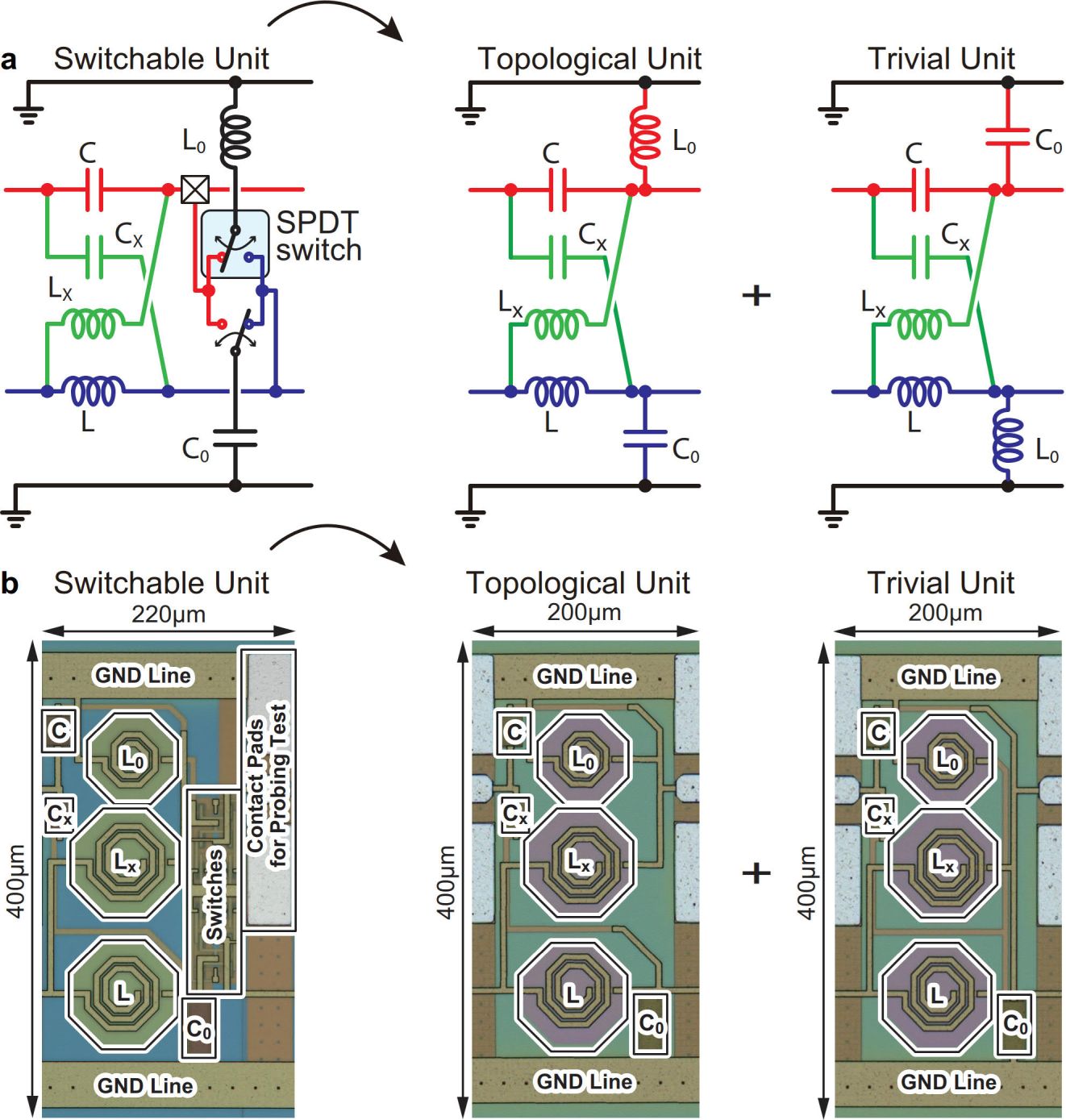}}
\caption{Switch-removed Kitaev chain
\textbf{a}, Illustrations of removing the switches from the Kitaev chain
\textbf{b}, A zoom-in view of removing the switches from the unit cell layout.}
\label{1stchip_diagram}
\end{figure}

We used SPDT switches with transistors to control the topological properties of each unit cell in the 16-unit circuit chain. However, these switches introduce parasitic capacitance that impacts the behavior of the Kitaev circuit chains.
To verify the impact of the switches on the Kitaev chain behavior, we have also designed Kitaev chains without switches. As shown in Supplementary  Fig.~\ref{1stchip_diagram}a, the SPDT switches are removed from the unit cell, and two different unit cells for topological and trivial phases are realized by replacing the connections of $L_0$ and $C_0$ with metal wires.


\begin{figure}[h]
\centerline{\includegraphics[width=0.9\textwidth]{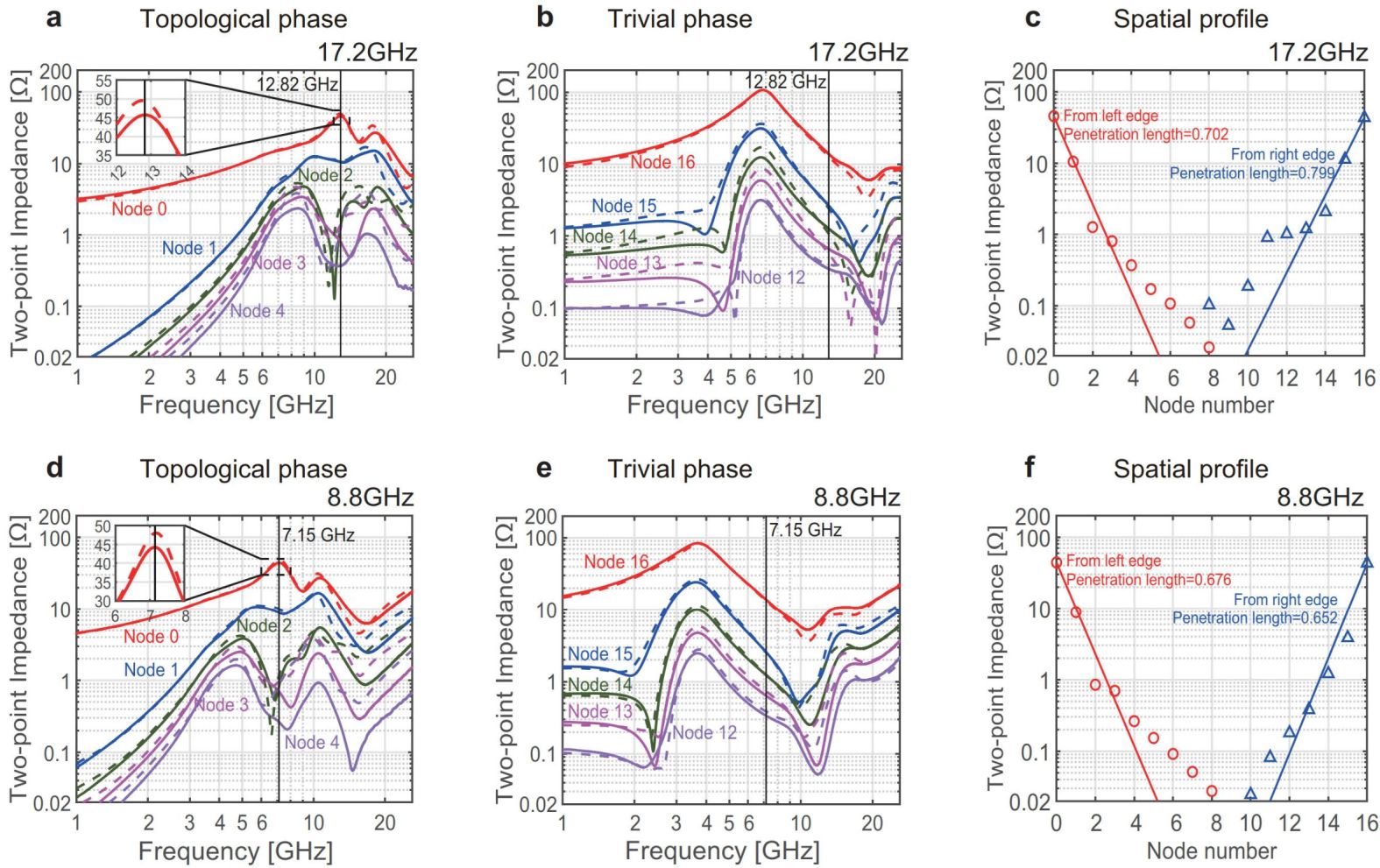}}
\caption{
\textbf{a}, 
Frequency dependence of the impedance parameters measured from the left edge of the \red{12.8\,GHz} Kitaev chain for the topological-fixed phase.
\textbf{b},
Frequency dependence of the impedance parameters measured from the right edge of the \red{12.8\,GHz} Kitaev chain for the trivial-fixed phase.
Solid and dashed lines show the measurement and simulation results of the Kitaev chains, respectively.
\textbf{c}, The spatial profile of the impedance values for the topological phase circuit chain measured from both left and right edges at their measured resonant frequencies.
\textbf{d}, \textbf{e} and \textbf{f}, Similar results for the circuit with 
\red{the resonant frequency of 7.2\,GHz}.
\red{The solid and dashed curves show measurement and simulation results, respectively.}
}
\label{FigResultsSI}
\end{figure}

Supplementary  Figs.~\ref{FigResultsSI}a, b and c summarize the impedance measurement results of the Kitaev chain without switches designed for \red{12.8\,GHz} resonant frequency. 
Supplementary  Figs.~\ref{FigResultsSI}a and b show the frequency dependence of the impedance measured from the left edge of the chain for topological and right edge of the chain for trivial, respectively.
The solid and dashed lines show measurement and simulation results, respectively.

In Supplementary  Fig.~\ref{FigResultsSI}a, an impedance peak is clearly observed from the leftmost edge of the topological chain. \red{The measured resonant frequency is 12.82\,GHz while the resonant frequency directly calculated based on the on-chip inductance and capacitance values is 17.2\,GHz. This resonant frequency shift is mainly caused by the parasitic inductance of the metal wires in the unit cell to connect the circuit elements. Since the parasitic impact is inevitable on the integrated chip, we utilized EM simulation to tune the actual resonant frequency.} On the other hand, there is no impedance peak observed from the rightmost edge of the trivial chain as shown in Supplementary  Fig.~\ref{FigResultsSI}b. For both topological and trivial chains, the measurement and simulation results agree almost perfectly, which indicates that the discrepancy observed for the chain with switches is caused mainly by the switch transistors.

Supplementary  Fig.~\ref{FigResultsSI}c summarizes the two-point impedance values at their measured resonant frequencies. The blue and red lines show the impedance measured from the left and the right edges, respectively. The leftmost (0-th) and rightmost (16-th) node impedance correspond to $Z_{11}$ and $Z_{22}$ values of the $2\times2$ impedance matrix.
The penetration length of the topological edge state is 0.702 unit cell for the left edge and 0.799 unit cell for the right edge. 
Compared with the measurement results for the chain with switches, these values show better agreement with the theoretical value 0.610 unit cell.


Supplementary  Figs.~\ref{FigResultsSI}d, e and f summarize the impedance measurement results of the Kitaev chain without switches designed for \red{7.2\,GHz} resonant frequency. 
Supplementary  Figs.~\ref{FigResultsSI}d and e show the frequency dependence of the impedance measured from the left edge of the chain for topological and right edge of the chain for trivial, respectively. 
The solid and dashed lines show measurement and simulation results, respectively. 
Similarly to the results for \red{12.8\,GHz} chain, an impedance peak is clearly observed from the leftmost edge of the topological chain as shown in Supplementary  Fig.~\ref{FigResultsSI}d. 
\red{The resonant frequency directly calculated based on the on-chip inductance and capacitance values is 8.8 GHz. 
Again, this is due to the effect of the parasitics which is inevitable on the integrated chip. For precise estimation of the actual resonant frequency including the impact of wirings, we utilized the EM simulation.} 
There is no impedance peak observed from the rightmost edge of the trivial chain as shown in Supplementary  Fig.~\ref{FigResultsSI}e. The measurement and simulation results agree almost perfectly also for this case, which again indicates that the discrepancy observed for the chain with switches is caused mainly by the switch transistors.

Supplementary  Fig.~\ref{FigResultsSI}f summarizes the two-point impedance values at their measured resonant frequencies. The blue and red lines show the impedance measured from the left and the right edges, respectively. The leftmost (0-th) and rightmost (16-th) node impedance correspond to $Z_{11}$ and $Z_{22}$ values of the $2\times2$ impedance matrix.
The penetration length of the topological edge state is 0.676 unit cell for the left edge and 0.652 unit cell for the right edge, which show a good agreement with the theoretical value 0.680 unit cell. 

\section{Supplementary Note 2. Topological edge state of the Kitaev Hamiltonian}

The Kitaev Hamiltonian reads%
\begin{equation}
H(k)=\frac{1}{2}\left( 
\begin{array}{cc}
-t\cos k-\mu  & i\Delta _{0}\sin k \\ 
-i\Delta _{0}\sin k & t\cos k+\mu 
\end{array}
\right) .
\end{equation}
We make a unitary transformation%
\begin{eqnarray}
H_{2}\left( k\right)  &=&UH(k)U^{-1}=\frac{1}{2}\left( 
\begin{array}{cc}
0 & -t\cos k-\mu -i\Delta _{0}\sin k \\ 
-t\cos k-\mu +i\Delta _{0}\sin k & 0
\end{array}
\right)  \\
&=&\frac{1}{2}\left( 
\begin{array}{cc}
0 & -t\frac{e^{ik}+e^{-ik}}{2}-\mu -i\Delta _{0}\frac{e^{ik}-e^{-ik}}{2i} \\ 
-t\frac{e^{ik}+e^{-ik}}{2}-\mu +i\Delta _{0}\frac{e^{ik}-e^{-ik}}{2i} & 0
\end{array}
\right)  \\
&=&\frac{1}{4}\left( 
\begin{array}{cc}
0 & -\left( t+\Delta _{0}\right) e^{ik}-\left( t-\Delta _{0}\right)
e^{-ik}-2\mu  \\ 
-\left( t-\Delta _{0}\right) e^{ik}-\left( t+\Delta _{0}\right) e^{-ik}-2\mu 
& 0
\end{array}
\right) 
\end{eqnarray}
with
\begin{equation}
U=\frac{1}{\sqrt{2}}\left( 
\begin{array}{cc}
1 & 1 \\ 
1 & -1
\end{array}
\right) ,
\end{equation}
We further make a unitary transformation
\begin{equation}
H_{3}\left( k\right) =U_{2}H(k)U_{2}^{-1}=\left( 
\begin{array}{cc}
0 & -\left( t+\Delta _{0}\right) -\left( t-\Delta _{0}\right) e^{-2ik}-2\mu
e^{-ik} \\ 
-\left( t+\Delta _{0}\right) -\left( t-\Delta _{0}\right) e^{2ik}-2\mu e^{ik}
& 0
\end{array}
\right) 
\end{equation}
with
\begin{equation}
U_{2}=\left( 
\begin{array}{cc}
1 & 0 \\ 
0 & e^{ik}
\end{array}
\right) .
\end{equation}
If we set
\begin{equation}
t_{A}=-\left( t+\Delta _{0}\right) ,\qquad t_{B}=-\left( t-\Delta
_{0}\right) ,\qquad \mu =0,\qquad k^{\prime }=2k,
\end{equation}
it is identical to the SSH Hamiltonian
\begin{equation}
H_{3}\left( k\right) =\left( 
\begin{array}{cc}
0 & t_{A}+t_{B}e^{-ik} \\ 
t_{A}+t_{B}e^{ik} & 0
\end{array}
\right) .
\end{equation}
Hence, the spatial profile of the topological edge state is identical to that of the SSH model.
The penetration depth of the Kitaev Hamiltonian is given by
\begin{equation}
\zeta =\frac{1}{\log  \left| t_{B}/t_{A}\right|  }=\frac{1}{\log \left| (t-\Delta _{0})/(t+\Delta _{0})\right| }.
\end{equation}
There is a correspondence
\begin{equation}
\left\{\begin{array}{l}
t=-C, \\
\mu=-2 C+C_0, \\
\Delta_0=-C_x.
\end{array}\right.
\end{equation}
The penetration length in terms of the capacitance reads
\begin{equation}
\zeta_{\text{SSH}} =\frac{1}{\log \left| (C+C_x)/(C-C_x)\right| }.
\end{equation}
We note that the actual penetration length of the Kitaev chain $\zeta_{\text{Kitaev}}$  is twice of that $\zeta_{\text{SSH}}$  of the SSH chain because there are two sites horizontally in the SSH model and vertically in the Kitaev model.

The penetration length is $\zeta_{\text{Kitaev}} =2\zeta_{\text{SSH}} =0.610$ for the chain with $C=220$fF and $C_x=204$fF, whose resonant frequency is $17.2$GHz.
The measured penetration length for the Kitaev chain without switches is $0.702$ for the left edge and $0.799$ for the right edge, where the measured resonant frequency is $12.68$GHz.
For the chain with switches, the measured penetration length is $0.860$ for the left edge and $0.788$ for the right edge, where the measured resonant frequency is $9.51$GHz.

The penetration length is $\zeta_{\text{Kitaev}} =2\zeta_{\text{SSH}}  =0.680$  with $C=440$fF and $C_x=396$fF, whose resonant frequency is $8.8$GHz.
The measured penetration length for the Kitaev chain without switches is $0.676$ for the left edge and $0.652$ for the right edge, where the measured resonant frequency is $7.08$GHz.
For the chain with switches, the measured penetration length is $0.771$ for the left edge and $0.916$ for the right edge, where the measured resonant frequency is $7.75$GHz.

Since the switches designed with transistors introduces parasitic capacitance that leads to less sharp resonance peak in the measurement results, the penetration lengths measured for the Kitaev chains with SPDT switches show more discrepancy from the theoretical values, while the measured values for the chains without switches exhibit reasonable agreement with the theoretical values.

\section{Supplementary Note 3. Effects of finite chain length}

\red{There are discrepancies for small impedance in Fig.2c and f in the main text. A possible reason would be the hybridization effect of two topological edge states for a finite length chain. To examine this possibility, we modify the fitting function as
\begin{equation}
Z_n\propto e^{-n/\zeta}+e^{(n-L)/\zeta}.
\end{equation}%
We show the fitting in Supplementary  Fig.\ref{FigCosh}. However, the fitting is not so good, indicating that the deviation from the exponential fit is not the hybridization effect but caused by the SPDT switches designed with transistors.}

\begin{figure}[h]
\centerline{\includegraphics[width=0.88\textwidth]{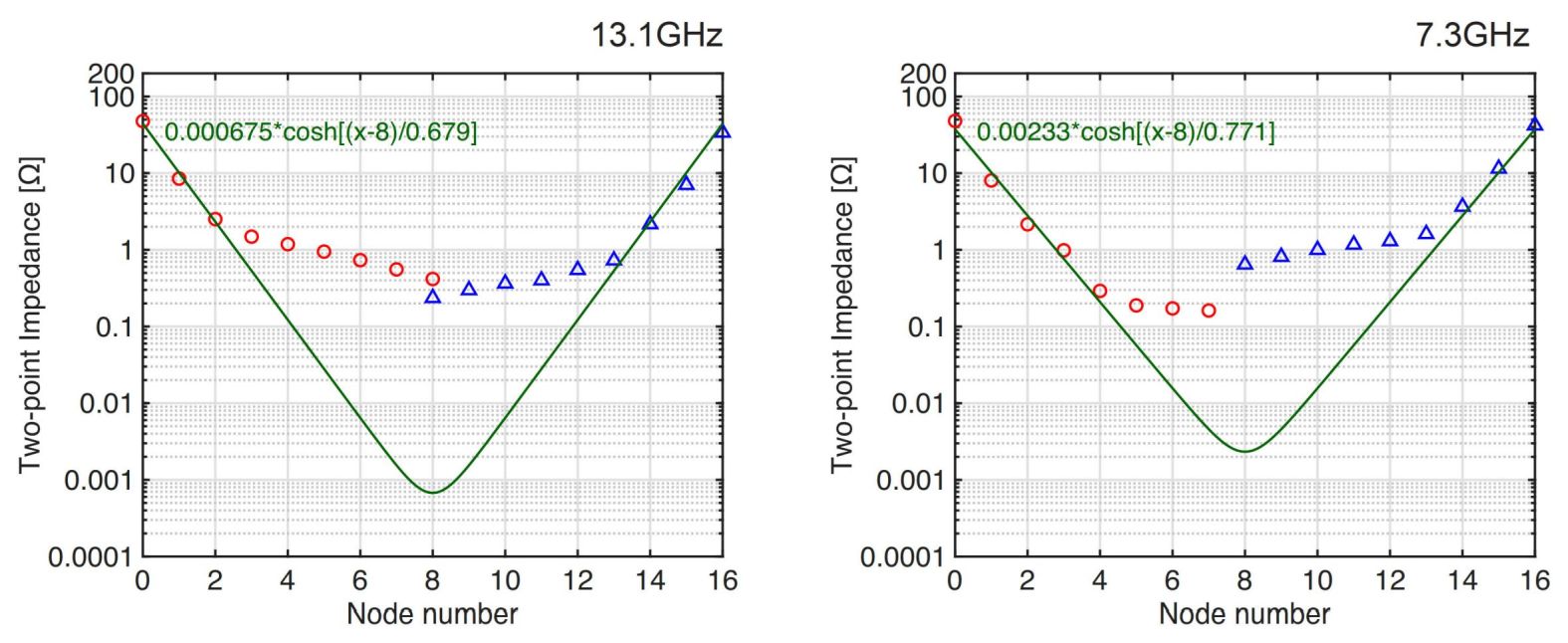}}
\caption{\red{The cosh-type fitting curves are shown in green curves. See the captions of Fig.2c and f in the main text.}}
\label{FigCosh}
\end{figure}

\section{Supplementary Note 4. Parasitic inductance}

\begin{figure}[h]
\centerline{\includegraphics[width=0.88\textwidth]{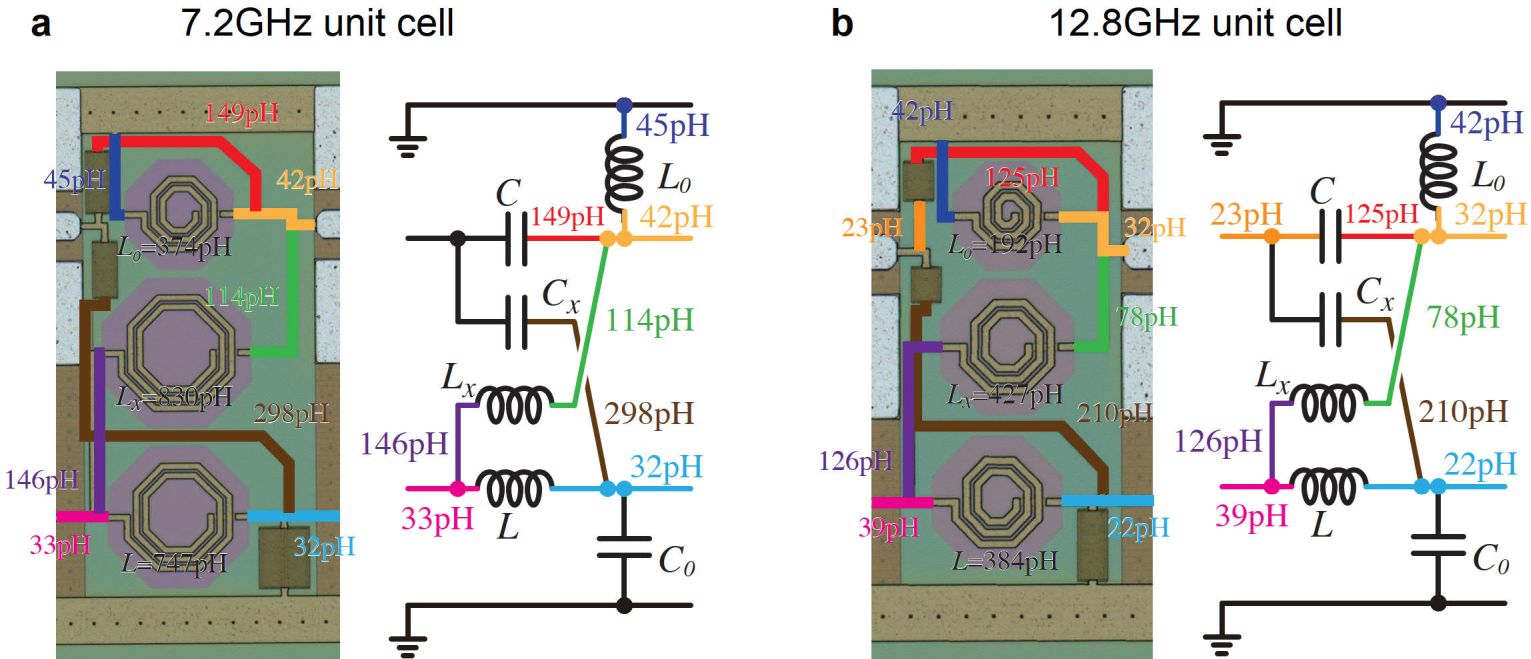}}
\caption{Parasitic inductance extraction results with EM simulation for \textbf{a} 8.8\,GHz unit cell and \textbf{b} 17.2\,GHz unit cell.}
\label{FigParasite}
\end{figure}

The measured resonant frequencies for both \red{the high and low frequency chains} are shifted down from \red{the values calculated directly from the inductance and capacitance values} mainly due to the parasitic inductance of the metal wires in the unit cell to connect the circuit elements. We conducted the electro-magnetic (EM) simulations to extract the inductance of each wiring piece as shown in Supplementary  Figs.\ref{FigParasite}a and b for \red{7.2\,GHz and 12.8\,GHz} unit cells, respectively.

By taking into account the parasitic inductance series with $L_x$ for example, for \red{7.2\,GHz} chain $L_{x,para}$ including the parasitic components is 830\,pH $+$ 114\,pH $+$ 146\,pH$=$ 1090\,pH, hence the resonant frequency is given by $1/\left(2\pi\sqrt{L_{x,para} C_x}\right)=7.7$\,GHz, which is much closer to the measured resonant frequency \red{7.2\,GHz}. For \red{12.8\,GHz} chain, $L_{x,para}=$427\,pH $+$ 78\,pH $+$ 126\,pH $=$ 631\,pH, hence the resonant frequency is 14.0\,GHz, which is close to the measured value \red{12.8\,GHz}.
For the higher frequency chain, as the original inductance values is small, the impact of the wiring parasitic is more obvious.

\section{Supplementary Note 5. Effects of randomness}

\red{We study effects of randomness in the circuit elements. We introduce randomness into inductors and capacitors uniformly distributing from $-\xi $ to $\xi $. As a result the inductance $L$\ and the capacitance $C$ becomes node-dependent, $%
L\rightarrow L_{n}$ and $C\rightarrow C_{n}$, where%
\begin{equation}
L_{n}=L\left( 1+\eta _{n}\xi \right) , \\
C_{n}=C\left( 1+\eta _{n}\xi \right) ,
\end{equation}%
with $\eta _{n}$ being a random variable ranging from $-1$ to $1$. We obtain the energy spectrum of the Kitaev model in the presence of randomness, which is shown in Supplementary  Fig.\ref{FigDisorder}. In-gap states are separated from the bulk bands for all range $\xi <1$. Especially, the edge states remain almost at the zero energy for $\xi <0.2$. In our sample, the $\xi $ is of the order of $\xi\sim 0.01$. Hence, the topological edge states are practically robust in experiments.}

\begin{figure}[h]
\centerline{\includegraphics[width=0.48\textwidth]{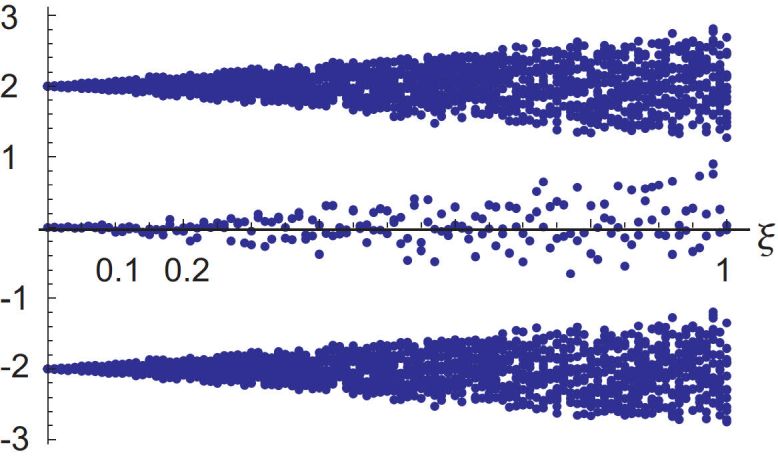}}
\caption{\red{Energy spectrum of the Kitaev model in the presence of randomness. The horizontal axis is the randomness, while the vertical axis is the energy.}}
\label{FigDisorder}
\end{figure}
\end{document}